\documentclass[
10pt, a4paper]{article}		
\usepackage{titlesec}
\usepackage{grffile}
\usepackage{authblk}		
\usepackage[centering,hmargin=15mm,vmargin=2cm]{geometry}
\usepackage[labelfont=bf]{caption} 
\usepackage[pdftex]{graphicx, xcolor}
\usepackage[backend=bibtex, autocite=superscript, style=nature, natbib=true]{biblatex}
\usepackage{bm} 
\usepackage{setspace}		
\usepackage{amsmath,amssymb}
\usepackage{xcolor}         
\usepackage{soul}             
\usepackage{wasysym}
\usepackage[normalem]{ulem} 
\usepackage[utf8]{inputenc}	
\usepackage{soul}
\usepackage[colorinlistoftodos]{todonotes}
\graphicspath{{./}}
\title{Softness suppresses fivefold symmetry and enhances crystallization of binary Laves phases in nearly hard spheres}

\author[1,$^\|$]{Tonnishtha Dasgupta}
\author[1,$^\|$]{Gabriele M. Coli}
\author[1,*]{Marjolein Dijkstra}

\affil[1]{Soft Condensed Matter, Debye Institute for Nanomaterials Science, Utrecht University, Princetonplein 5, 3584 CC Utrecht, The Netherlands.}
\affil[$^\|$]{These authors contributed equally to this work.}
\affil[*]{e-mail: m.dijkstra@uu.nl}

\date{}
\bibliography{citation}
\nocite{*}
\begin{document}

\maketitle

\noindent \textbf{Colloidal crystals with a diamond and pyrochlore structure display wide photonic band gaps at low refractive index contrasts. However, these low-coordinated and open structures are notoriously difficult to self-assemble from colloids interacting with simple pair interactions.  To circumvent these problems, one can self-assemble both structures in a closely packed MgCu$_2$ Laves phase from a binary mixture of colloidal spheres and then selectively remove one of the sublattices. Although Laves phases have been proven to be  stable in a binary hard-sphere system, they have never been observed to spontaneously crystallize in such a fluid mixture in  simulations nor in experiments of  micron-sized hard spheres due to slow dynamics. Here we demonstrate, using computer simulations, that softness in the interparticle potential suppresses  the degree of fivefold symmetry  in the binary fluid phase and enhances crystallization of Laves phases in nearly hard spheres. 
}

\bigskip

Photonic crystals (PCs) are periodic dielectric structures that possess a photonic bandgap that forbids the propagation of light at  certain frequency ranges.  The ability to control the flow of light opens the way to numerous applications, ranging from  lossless dielectric mirrors, bending of light around sharp corners in optical waveguides, telecommunications, to optical transistors in optical computers. A highly promising route to fabricate photonic crystals is \emph{via} self-assembly of  optical wavelength sized colloidal building blocks. 
PCs that display a wide omnidirectional photonic bandgap at low refractive index contrasts are related to the family of either the diamond or the pyrochlore structure. However, these  low-coordinated crystals are notoriously difficult to self-assemble from colloids with simple isotropic pair interactions. One strategy to form open lattices is by employing long-range Coulomb interactions with a range that exceeds multiple times the particle size \autocite{kalsin2006electrostatic,bishop2013and}. The range of the screened Coulomb interaction  is set by the Debye screening length of the solvent, like water or other polar solvents, which is why this approach will fail  for particle sizes that are required for opening up a photonic bandgap in the visible region. 

\begin{figure}
\includegraphics[width=1.0\textwidth, keepaspectratio]{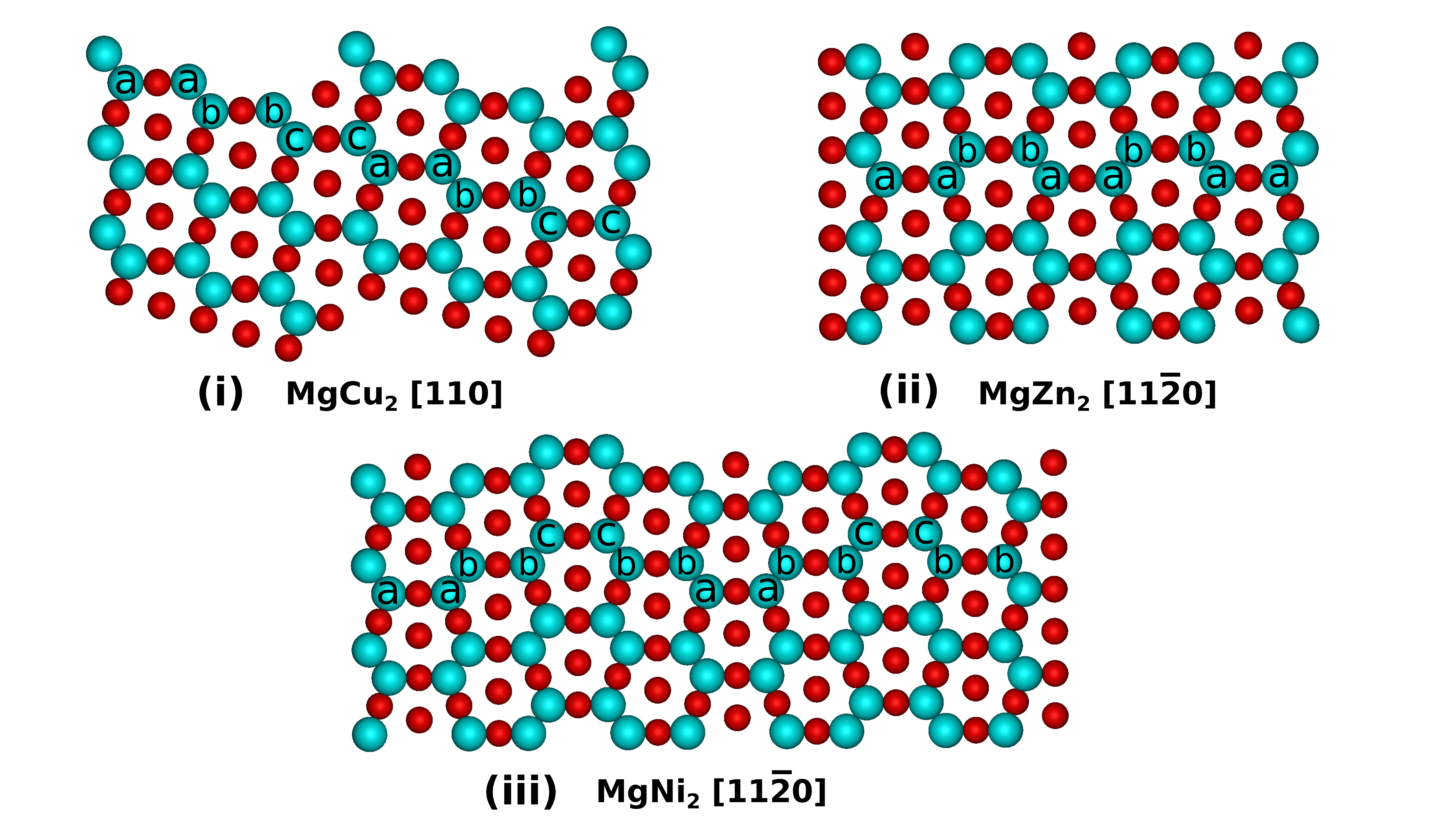}
\caption{Structure of the three  types of Laves phases, showing the different stacking sequences of the large-sphere dimers, marked as "aa", "bb" and "cc", when viewed along specific projection planes. The stacking of the large-sphere dimers is (i) ``...aa-bb-cc...'' for MgCu$_2$, (ii) ``...aa-bb...'' for MgZn$_2$, and (iii) ``...aa-bb-cc-bb...'' for MgNi$_2$.  }
\label{LPtype}
\end{figure}

\noindent To circumvent these problems associated with the self-assembly of  low-coordinated crystal structures, one can also employ a different route in which both the diamond and pyrochlore structure are self-assembled in a single close-packed MgCu$_2$ crystal structure from a binary colloidal dispersion. By selectively removing one of the species, one can obtain either the diamond  (Mg, large spheres) or the pyrochlore (Cu, small spheres) structure. MgCu$_2$ is one of the three \emph{binary}  $LS_2$ crystal structures  ($L$ = large species, $S$ = small species), also known as Laves phases (LPs), as first found in intermetallic compounds.  The three main structural prototypes of the  LPs are the hexagonal MgZn$_2$, cubic MgCu$_2$ and hexagonal MgNi$_2$ structures, which can be  distinguished by the stacking of the large-sphere dimers in the crystal structures (Fig.~\ref{LPtype}). \noindent Experimentally, LPs have been observed in binary nanoparticle suspensions \autocite{shevchenko2006structural,evers2010entropy}, and in submicron-sized spheres interacting \emph{via}  soft repulsive  potentials  \autocite{yoshimura1983order,hasaka1984structure,ma1994preparation,gauthier2004phase,cabane2016hiding,schaertl2018formation}.

\noindent Although free-energy calculations in Monte Carlo (MC) simulations have demonstrated that the LPs are thermodynamically stable for a  binary hard-sphere (BHS) mixture with a diameter ratio of $0.76 \leq q = \sigma_S/\sigma_L \leq 0.84$ \autocite{hynninen2007self}, LPs have never been observed to spontaneously crystallize in such a binary fluid mixture in computer simulations. There are numerous possible reasons. First of all, it may be possible that the LPs are not stable in such a hard-sphere mixture and should be replaced by a crystal structure that has been ignored  so far in  phase diagram calculations  \autocite{hynninen2007self}. Secondly, the freezing transition of the LPs in a BHS fluid  is located at very high densities. Nucleation can thus only occur when the system is  sufficiently dense. At these high concentrations,  nucleation is severely hampered by slow dynamics. Binary mixtures with a diameter ratio of $q \sim 0.8$, identical to the range where the LPs are stable, are known to be excellent glassformers \autocite{jungblut2011crystallization}. Furthermore, due to small free-energy differences, the three LPs are strongly competing  during the crystallization process, which in conjunction with the above factors makes the self-assembly of LPs in BHS mixtures an extremely rare event. The suppression of crystallization due to glassy behaviour is  often rationalized by the prevalence of  icosahedral clusters of spheres  whose short-range fivefold  symmetry is  incompatible with the long-range translational order as exhibited by crystals \autocite{frank1952supercooling}. The icosahedral order arises  when one maximises the density, using the convex hull, of a packing of 12 identical spheres in contact with a central sphere of the same size. The densest packing is obtained by arranging the outer spheres on the vertices of an icosahedron, rather than by using 13-sphere subunits of face-centered cubic and hexagonal close-packed bulk crystals. 

Here we demonstrate that spontaneous crystallization of the  LPs  is strongly suppressed by the presence of fivefold symmetry structures in a binary fluid  of hard spheres. Interestingly, we show that softness of the interaction potential  reduces the degree of fivefold symmetry in the binary fluid phase. We systematically  study the role of softness in the interaction potential on the structure, phase behaviour, and nucleation of the LPs. 
By carefully tuning the particle softness, we  observe for the first time spontaneous nucleation of the LPs in a nearly hard-sphere system in computer simulations, thereby providing evidence  that the LPs are stable in a binary hard-sphere system. The key result of this study is that  soft repulsive spheres can be mapped onto a hard-sphere system in such a  way that the structure and thermodynamics are invariant, but that the dynamics and therefore the kinetic glass transition are strongly affected by higher-body correlations, \emph{i.e.} fivefold symmetry clusters, which can be tuned both in simulations and in experiments by the softness of the particle interactions. In this way, softness suppresses fivefold symmetry and enhances crystallization of the LPs. 

\section*{Results}
\subsection*{Freezing transition and Fivefold Symmetry}
\begin{figure}[t!]
\centering
\includegraphics[width=0.9\textwidth, keepaspectratio]{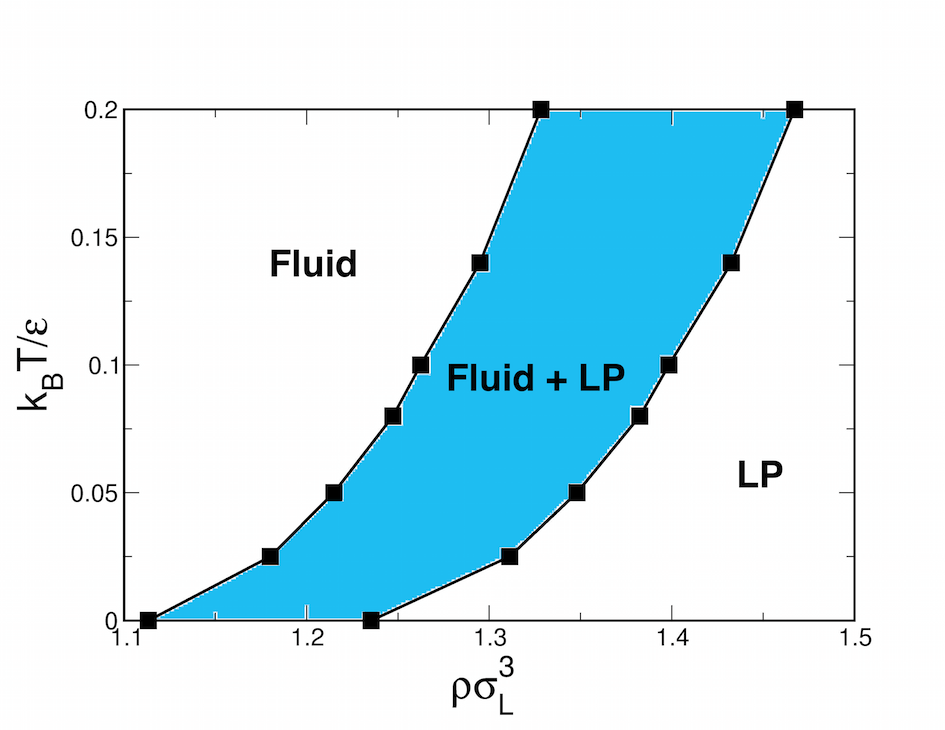}
\caption{Fluid-Laves Phase (LP) coexistence  as denoted by the blue region of a binary mixture of WCA spheres with a diameter ratio $q$ = 0.78 at a fixed composition $x_L =  N_L/(N_L+N_S)=1/3$ in the reduced temperature $k_B T / \epsilon$ - reduced density $\rho \sigma_L^3$ plane. In the limit of $k_BT/\epsilon \rightarrow 0$, the system reduces to a  binary  mixture of hard spheres.}
\label{Trho_iso}
\end{figure}

\begin{figure}[t!]
\centering
\includegraphics[width=0.9\textwidth, keepaspectratio]{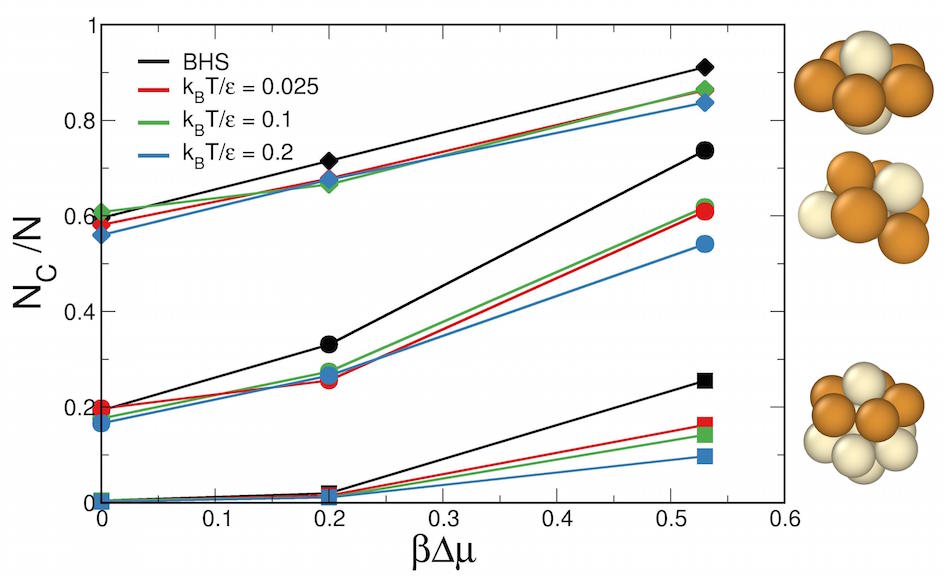}
\caption{Number fraction  of particles $N_c/N$ belonging to three different fivefold symmetry clusters as a function of the supersaturation $\beta \Delta \mu$ of the fluid phase of a binary mixture of WCA spheres at varying temperatures as labeled corresponding to different degrees of particle softness, and of a binary hard-sphere mixture. The three data sets correspond to pentagonal bipyramids (diamonds), defective icosahedra (bullets) and icosahedra (squares). Sketches of these clusters are shown on the right. We highlight (one of the) pentagons in the respective clusters. }
\label{clusters}
\end{figure}

We first study the effect of particle softness on the freezing transition of the LPs in a binary fluid of soft repulsive spheres, modelled by the Weeks-Chandler-Andersen (WCA) potential\autocite{weeks1971role}. The \emph{softness} of the potential can be tuned by changing the reduced temperature $T^*=k_BT/\epsilon$ with  $k_B$ Boltzmann's constant and $T$ the temperature. The WCA potential has been used previously to mimic the interactions between hard spheres \autocite{kawasaki2010formation,filion2011simulation,speck2018a,speck2018b}.
Using free-energy calculations and MC simulations we determine the fluid-LP coexistence for a binary WCA mixture at a composition $x_L = N_L/(N_L + N_S) = 1/3$, corresponding to the stoichiometry of the LPs (see Methods for details). The phase diagram is shown in Fig. \ref{Trho_iso} in the reduced temperature $k_BT/\epsilon$ - reduced density $\rho \sigma_L^3$ plane, where $\rho=(N_L+N_S)/V$ denotes the density.
We find that the freezing transition moves to higher $\rho\sigma^3_L$ with increasing temperature or softness of the particle interaction. In the limit of $k_BT/\epsilon \rightarrow 0$, the system reduces to a   binary hard-sphere mixture. 
The bulk densities of the fluid-LP coexistence for  a BHS mixture  with a diameter ratio $q=0.78$ correspond to  packing fractions  $\eta_\textrm{BHS}^{(f)}=0.5356$ and $\eta_\textrm{BHS}^{(LP)}=0.5943$ for the  fluid and LP, respectively.  
As the freezing transition of the LPs is located at relatively high densities, crystallization is likely suppressed by slow dynamics. In the case of monodisperse spheres, glassy dynamics and suppression of crystallization are often linked to the presence of icosahedral clusters with fivefold symmetry in the supersaturated fluid, which is incompatible with the long-range periodic order of a crystal. To investigate whether or not fivefold symmetry structures suppress crystallization of the LPs in a binary fluid mixture at composition $x_L = 1/3$, we measure the number fraction of three significant representatives of the fivefold symmetry structures, i.e. the pentagonal bipyramids, defective icosahedra, and regular icosahedral clusters as depicted in Fig.~\ref{clusters}, using the topological cluster classification (TCC) \autocite{malins2013identification} for varying softness of the interparticle potential. The effect of the presence of these clusters on the kinetics and nucleation of monodisperse hard-sphere systems has already been investigated \autocite{taffs2016role,wood2018coupling}. In order to investigate the effect of particle softness, we compare the number fraction of these clusters  at  fixed supersaturation $\beta\Delta \mu$ for varying temperatures $T^*$. The supersaturation $\beta \Delta \mu=\beta\mu_{\textrm{\small fluid}}(P)-\beta \mu_{\textrm{\small LP}}(P)$ is defined as the   chemical potential difference  between the supersaturated fluid and the stable LP at pressure $P$.

\begin{figure}[t!]
\centering
\includegraphics[width=1.0\textwidth, keepaspectratio]{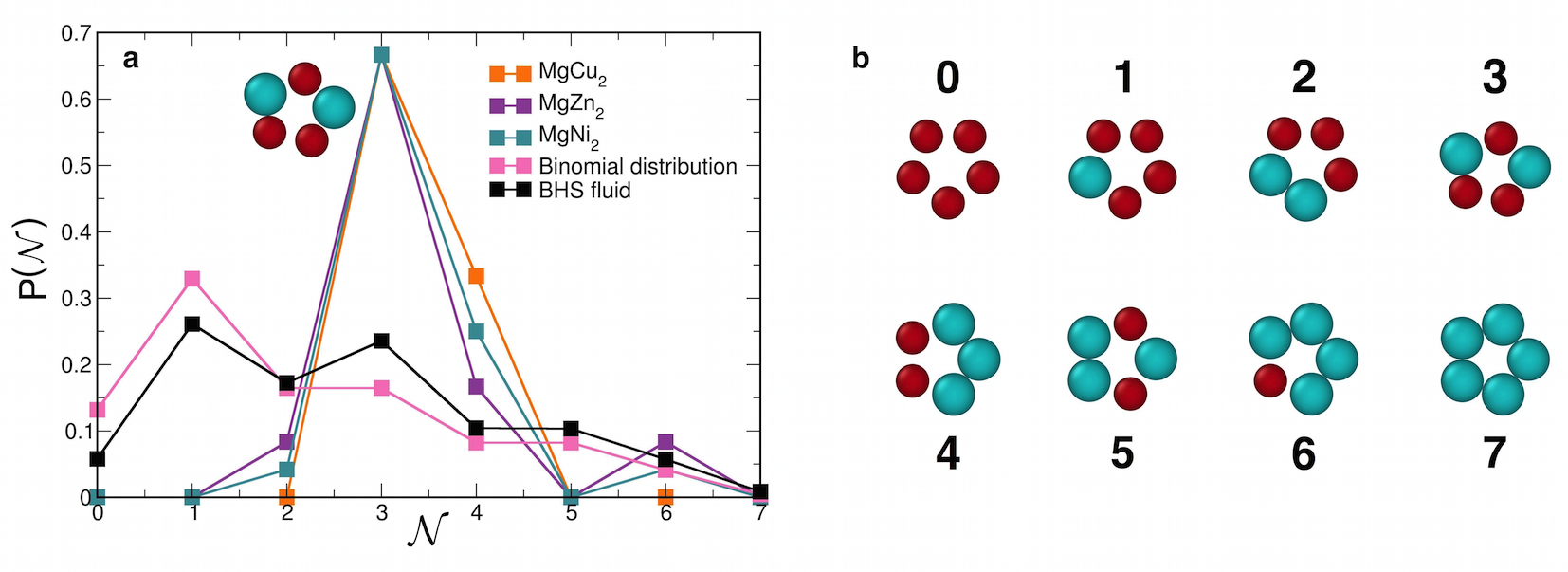}
\caption{a) Probability distribution to observe a specific cluster topology   $P({\cal N})$ for the five-membered rings for the three ideal LPs,  a supersaturated binary hard-sphere (BHS) fluid ($\beta\Delta \mu$ $\simeq$ 0.533), and  a binomial distribution, all at  composition $x_L = 1/3$. The 8 distinct cluster topologies are shown in b) with their index label ${\cal N}$. }
\label{pentagons}
\end{figure}

In Fig.~\ref{clusters}, we plot the number fraction $N_c/N$ of the three investigated clusters in a binary fluid mixture  at composition $x_L=1/3$  versus $\beta\Delta \mu$ for varying $T^*$ corresponding to different particle softness. We clearly see that the number fraction of fivefold symmetry clusters increases with $\beta\Delta \mu$, but more remarkably, it decreases substantially with a small increase in particle softness. 
Notably, the five-membered rings as observed in the clusters highlighted in Fig.~\ref{clusters} are also prevalent in the three LP crystal structures.  It is thus not immediately clear whether these pentagons in the supersaturated fluid act as nucleation precursors or are responsible for the slow dynamics. 
We therefore analyse the five-membered rings further in the BHS fluid phase as well as in the three ideal LPs,  and classify them according to their large/small sphere composition and topology. 
We distinguish 8 topologies  (indexed ${\cal N}$)  in Fig.~\ref{pentagons}b, and measure the probability to observe a specific topology  $P({\cal N})$ in the ideal LPs and the metastable BHS fluid at a high supersaturation $\beta\Delta \mu$ $\simeq$ 0.53. 
We reason that if these five-membered rings are formed randomly in a fluid mixture, $P({\cal N})$ should follow a binomial distribution where the probability to observe a large sphere in a pentagonal cluster is determined by the composition $x_L$.
We present the probability distributions $P({\cal N})$ for the LPs,  the metastable fluid, and  binomial distribution  all at a  composition  $x_L=1/3$  in Fig.~\ref{pentagons}a. We find that the probability distribution $P({\cal N})$ of the pentagons in the BHS fluid mixture (black line in Fig.~\ref{pentagons}a) follows reasonably well the binomial distribution (pink line in Fig.~\ref{pentagons}a) for all topologies,  demonstrating that the pentagons  are formed randomly in the fluid. Furthermore, the pentagons with a topology ${\cal N}=2$ are predominant in the supersaturated BHS fluid phase, whereas pentagons with a topology ${\cal N}=3$ and 4 are prevalent in the ideal LPs.
We therefore conclude   that  the  fivefold symmetry clusters in the supersaturated fluid do not act as precursors for crystallization, but are responsible for the slowing down of the dynamics and the kinetic arrest. Moreover, we find that the presence of these fivefold symmetry clusters can be reduced significantly by particle softness. This unexpected finding raises the immediate question whether or not crystal nucleation of the LPs can be enhanced or suppressed by tuning the softness of the interparticle potential. 

\subsection*{Nucleation behaviour} 
\begin{figure}[t!]
\centering
\includegraphics[width=0.9\textwidth, keepaspectratio]{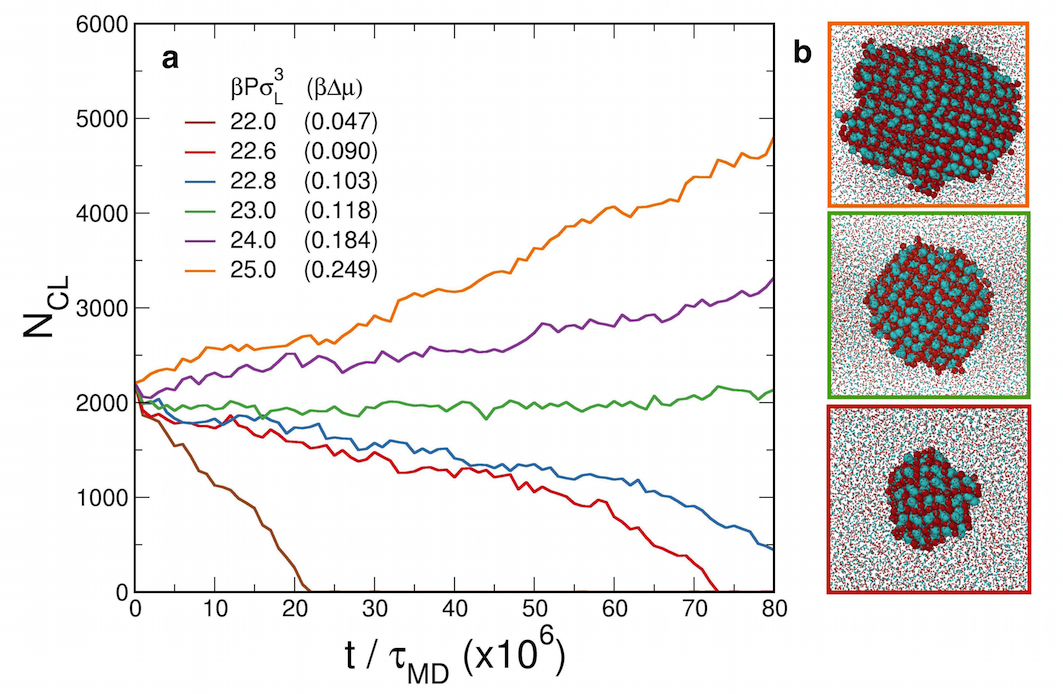}
\caption{(a) The largest cluster size $N_\textrm{\footnotesize CL}$ with LP symmetry as a function of time $t/\tau_\textrm{\small MD}$ using the seeding approach in MD simulations of a binary mixture of WCA spheres in the \emph{NPT} ensemble at temperature $T^*=0.2$, composition $x_L=1/3$ and a diameter ratio $q=0.78$ for varying pressures $\beta P \sigma_L^3$ with corresponding supersaturations $\beta \Delta \mu$ between brackets in order to estimate the critical pressure $\beta P_c \sigma_L^3$. The initial seed size is 2688 particles of the MgZn$_2$ Laves phase. The snapshots show (b) the melting of the seed at $\beta P \sigma_L^3=22.6$, (c) growth of the seed at $\beta P \sigma_L^3=25$, and (d) a more or less stable seed size at the critical pressure $\beta P_c \sigma_L^3=23$. The large (small)  spheres are coloured blue (red). Fluid particles are reduced in size for visual clarity.}
\label{PCS_seeding}
\end{figure}

\begin{figure}[t!]
\centering
\includegraphics[width=0.9\textwidth, keepaspectratio]{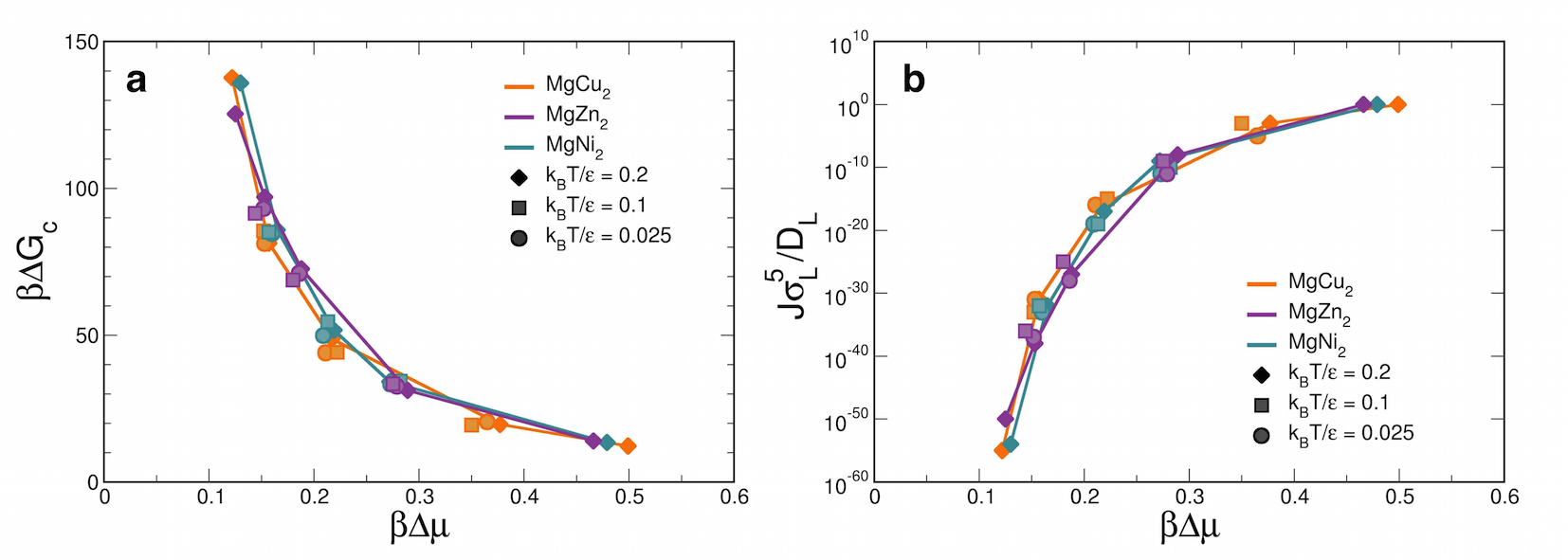}
\caption{a) The height of the Gibbs free-energy barrier $\beta \Delta G_c$  and  b) the nucleation rate $J \sigma_L^5 / D_L$ as a function of the chemical potential difference $\beta \Delta \mu$ between the fluid and the LP for a binary WCA mixture for the three different LPs and for temperatures $k_BT/\epsilon=0.025, 0.1$, and $0.2$.}
\label{nucl_barrier}
\end{figure}

To investigate the effect of particle softness on the nucleation of the LPs, we determine the nucleation barrier height, critical nucleus size, and nucleation rate using the seeding approach~\autocite{espinosa2016seeding,espinosa2017lattice}.  
This technique involves inserting a crystalline seed of a pre-determined shape and size into a metastable fluid. The configuration is subjected to a two-step equilibration process in the \emph{NPT} ensemble where (i) the interface between the crystalline cluster and surrounding fluid is equilibrated by keeping the cluster fixed and then (ii) the constraint on the cluster is relaxed and the system is equilibrated further. Subsequently, the equilibrated configuration is simulated for a range of pressures in order to determine the critical pressure $\beta P_c \sigma_L^3$ at which the critical cluster size of $N_c$ particles stabilizes. 
An illustration is shown in Fig.~\ref{PCS_seeding}, where a MgZn$_2$ LP seed melts, stabilizes and grows out, as can be observed from the evolution of the size of the largest cluster $N_\textrm{\footnotesize Cl}$ as a function of time $t/\tau_{MD}$, at $\beta P \sigma_L^3$ = 22.6 (red curve), 23 (green) and 25 (orange), respectively. Here, $\tau_{MD} = \sigma_L\sqrt{m/k_BT}$ denotes the MD time unit and $m$ the mass of the particles.
The bond order parameter criterion to distinguish LP clusters from fluid particles can be found in the Methods section.
The height of the Gibbs free-energy barrier $\Delta G_c$ for a critical nucleus size $N_c$ can subsequently be obtained from Classical Nucleation Theory
\begin{equation}
\beta \Delta G_c (N_c) = N_c ~\beta \Delta\mu/2. 
\label{gibbs}
\end{equation}
By using  different critical cluster sizes $N_c$ in the seeding approach, we obtain $\Delta G_c$ for varying critical pressures, corresponding to different supersaturations $\beta\Delta \mu$. We repeat these calculations for the three distinct LPs, MgZn$_2$, MgCu$_2$ and MgNi$_2$, as crystalline seeds in the seeding approach. In  Fig.~\ref{nucl_barrier}a, we present $\Delta G_c$ as a function of $\beta\Delta \mu$ for the three LPs and for temperatures $k_B T/\epsilon$ = 0.025, 0.1 and 0.2, corresponding to varying particle softness.  We observe that for all temperatures and the three LP types,  $\Delta G_c$  goes to infinity upon approaching bulk coexistence at $\beta \Delta \mu = 0$,  decreases with increasing  supersaturation $\beta\Delta \mu$, and approaches zero at sufficiently high $\beta\Delta \mu$. We find that 
all our $\Delta G_c$ data  coincides within statistical error bars  for all the three  LPs, which is to be expected as the free-energy differences between the three bulk LPs are extremely small. More remarkably, we  observe that the $\Delta G_c$ data collapses onto a master curve for all  three  temperatures, yielding an intriguing thermodynamic invariance for the different degrees of softness in the WCA interaction potential. 
In addition, we  calculate the nucleation rate $J$, which is determined by a thermodynamic term related to the Gibbs free-energy barrier $\beta \Delta G_c$ and a kinetic pre-factor 
\begin{equation}
\frac{J \sigma_L^5}{D_L}  = \sqrt{\frac{\beta \Delta\mu}{6 \pi N_c}}\frac{f^{+}\sigma_L^2}{D_L}\rho_f  \sigma_L^3\exp(-\beta \Delta G_c), 
\label{nucl_rate}
\end{equation}
where $f^{+} = \langle\left(N(t)-N_c\right)^2\rangle/t$ is the attachment rate of particles to the critical cluster,  $t$ the time, $\rho_f(\beta P_c \sigma_L^3)$ is the critical density of the fluid at the critical pressure, and $D_L$ is the long-time diffusion coefficient at the same $\rho_f$. The attachment rate $f^+$ is measured from 10 independent simulation trajectories at  the critical density $\rho_f$. We present the nucleation rates as a function of $\beta\Delta \mu$ in Fig.~\ref{nucl_barrier}b, and find that they collapse  for all three temperatures and three LPs  onto a master curve - in a similar way as we observed for the nucleation barriers in Fig.~\ref{nucl_barrier}a. This finding  can be rationalized by the fact that the nucleation rate  is predominantly determined by the thermodynamic term and simply echoes the thermodynamic invariance as observed for $\beta \Delta G_c$ for the three LPs and the three temperatures. Moreover, we find that not only does the  nucleation barrier   
 decrease with supersaturation $\beta \Delta \mu$, but more importantly it also allows us to pinpoint the  supersaturation range, where  the nucleation barrier of the LP becomes so low that spontaneous nucleation should occur. 

\subsection*{Spontaneous nucleation}
\begin{figure}[t!]
\centering
\hspace*{-0.1in}
\includegraphics[width=0.98\textwidth, keepaspectratio, angle=0]{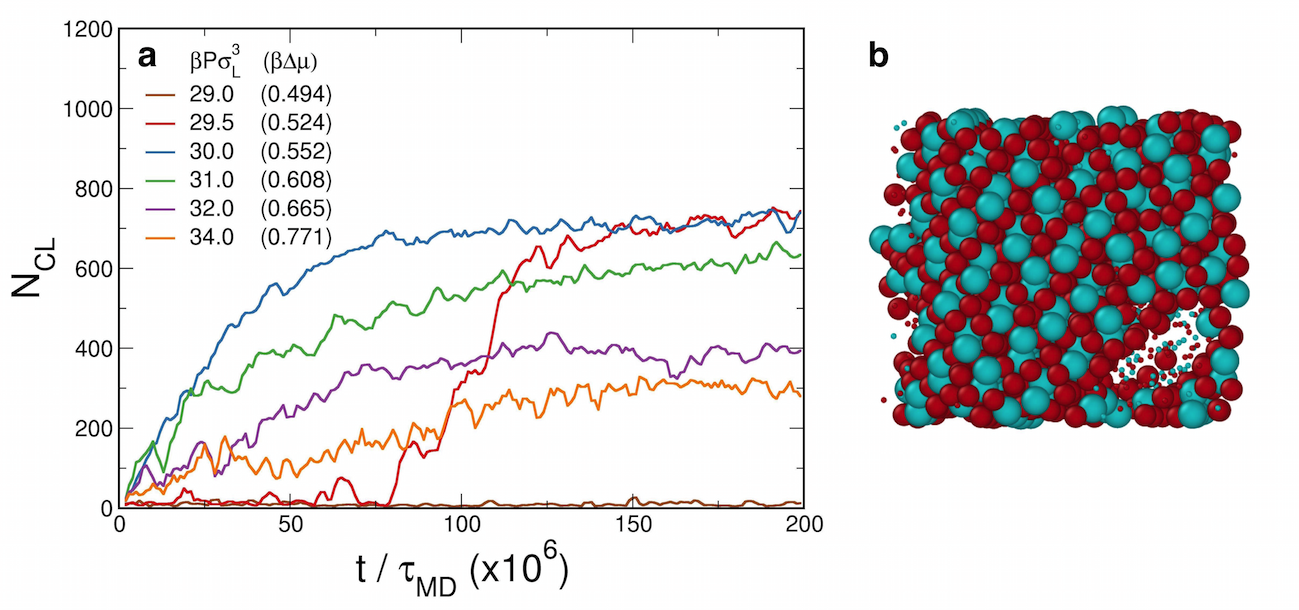}
\caption{a) Size of the largest crystalline cluster $N_\textrm{\footnotesize CL}$ for a binary mixture of WCA spheres with a diameter ratio $q = 0.78$ and temperature $T^*$ = 0.2 as a function of time $t/\tau_{MD}$ for varying pressures $\beta P\sigma_L^3$ with corresponding supersaturations $\beta \Delta \mu$ between brackets  using MD simulations in the \emph{NPT} ensemble. Profiles are averaged over three independent simulations. b) Configuration of the MgCu$_2$ phase,  spontaneously formed at   $\beta P\sigma_L^3=29.5$ ($\beta \Delta \mu=0.524$). }
\label{T0p2_PCS}
\end{figure}

Guided by the seeding approach results, we perform MD simulations in the \emph{NPT} ensemble and search for spontaneous nucleation of the LP  in a highly supersaturated binary fluid phase of soft repulsive spheres, which has hitherto never been observed in previous simulation studies. In Fig.~\ref{T0p2_PCS}, we determine the size of the largest cluster $N_\textrm{\footnotesize CL}$ with  LP symmetry as a function of  time $t/\tau_{MD}$ for a range of pressures beyond coexistence, for $k_BT/\epsilon$ = 0.2, where the coexistence pressure $\beta P \sigma_{L}^3$ = 21.32.  The results yield some interesting observations. At pressure $\beta P \sigma_{L}^3$ = 29, we find absence of crystallization within reasonable time scales of our  simulations. At a slightly higher pressure, $\beta P \sigma_{L}^3$ = 29.5 ($\beta \Delta \mu=0.524$), we observe that the system stays in a metastable fluid phase for a certain induction time until a nucleation event occurs, \emph{i.e.} a crystalline nucleus of the MgZn$_2$ phase forms that subsequently grows out  and transforms into the MgCu$_2$ phase as soon as the cluster spans the whole simulation box. Upon increasing the pressure further $\beta P \sigma_{L}^3 \geq 30$, the crystallization exhibits features of spinodal-like behaviour as the supersaturated fluid is unstable with respect to the crystal phase and small crystalline nuclei appear immediately throughout the system. For still higher pressures, we again  see immediate crystallization, but the clusters grow less, which we attribute to glassy behaviour. 
To study the effect of temperature or particle softness on the spontaneous nucleation of the LP in a binary mixture of WCA spheres, we perform MD simulations   in the \emph{NPT} ensemble for $T^* =$ 0.1 and 0.025 and pressures higher  than the coexistence pressures  $\beta P \sigma_{L}^3$ = 20.03 for $T^* = 0.1$, and $\beta P \sigma_{L}^3$ = 18.35 for $T^* = 0.025$,  respectively. We find a similar pressure-dependence (not shown) as described above for $T^*=0.2$: absence of crystallization at  low pressures, nucleation in a tiny intermediate pressure regime, and immediate spinodal-like crystallization at  $\beta P \sigma_{L}^3 \simeq$ 27.5 at  $T^* = 0.1$, and $\beta P \sigma_{L}^3 \simeq$ 25.5 for $T^*$ = 0.025.  Surprisingly, the values of  the thermodynamic driving force $\beta \Delta \mu$ corresponding to the pressures at which the fluid is unstable  are given by  $\beta\Delta\mu = 0.53 \pm 0.02$ for all three temperatures, which yields again an intriguing ``universality"  for the onset of spinodal-like behaviour.

\subsection*{Invariance with hard spheres}
Perturbation and integral equation theories of simple liquids are  based on the  premise that the structure of monatomic fluids at high densities resembles that of  hard spheres~\autocite{weeks1971role,hansen1990theory,rosenfeld1979theory}. Hence, a system of hard spheres serves as a natural reference system for determining the properties of more realistic systems. On this basis one  expects invariance of the structure along the melting and freezing line of simple fluids, and of thermodynamic properties such as the relative density change upon freezing and melting~\autocite{ross1969generalized,pedersen2016thermodynamics}. 
Here, we observe an  {\em invariance} of the Gibbs free-energy  barriers,  nucleation rates, and the onset of spinodal-like behaviour as a function of $\beta\Delta\mu$  for  binary  WCA mixtures at different temperatures. Inspired by this remarkable observation, we investigate whether other  thermodynamic quantities, {\em and} structural  properties are invariant along the freezing line of our WCA systems. Such an invariance is very interesting as it allows us to map the WCA mixture onto a simple binary hard-sphere system. A BHS mixture with a fixed composition depends on only one thermodynamic variable, the overall packing fraction, thereby  yielding a simple one-dimensional phase diagram with a unique freezing and melting transition. In addition, the invariance may enable us to make predictions on the nucleation of the LP in a binary hard-sphere mixture, and may shed  light on why LP nucleation is observed in a binary mixture of soft repulsive spheres and not in a system of hard spheres.


\begin{figure}[t!]
\centering
\hspace*{-0.1in}
\includegraphics[width=0.9\textwidth, keepaspectratio, angle=0]{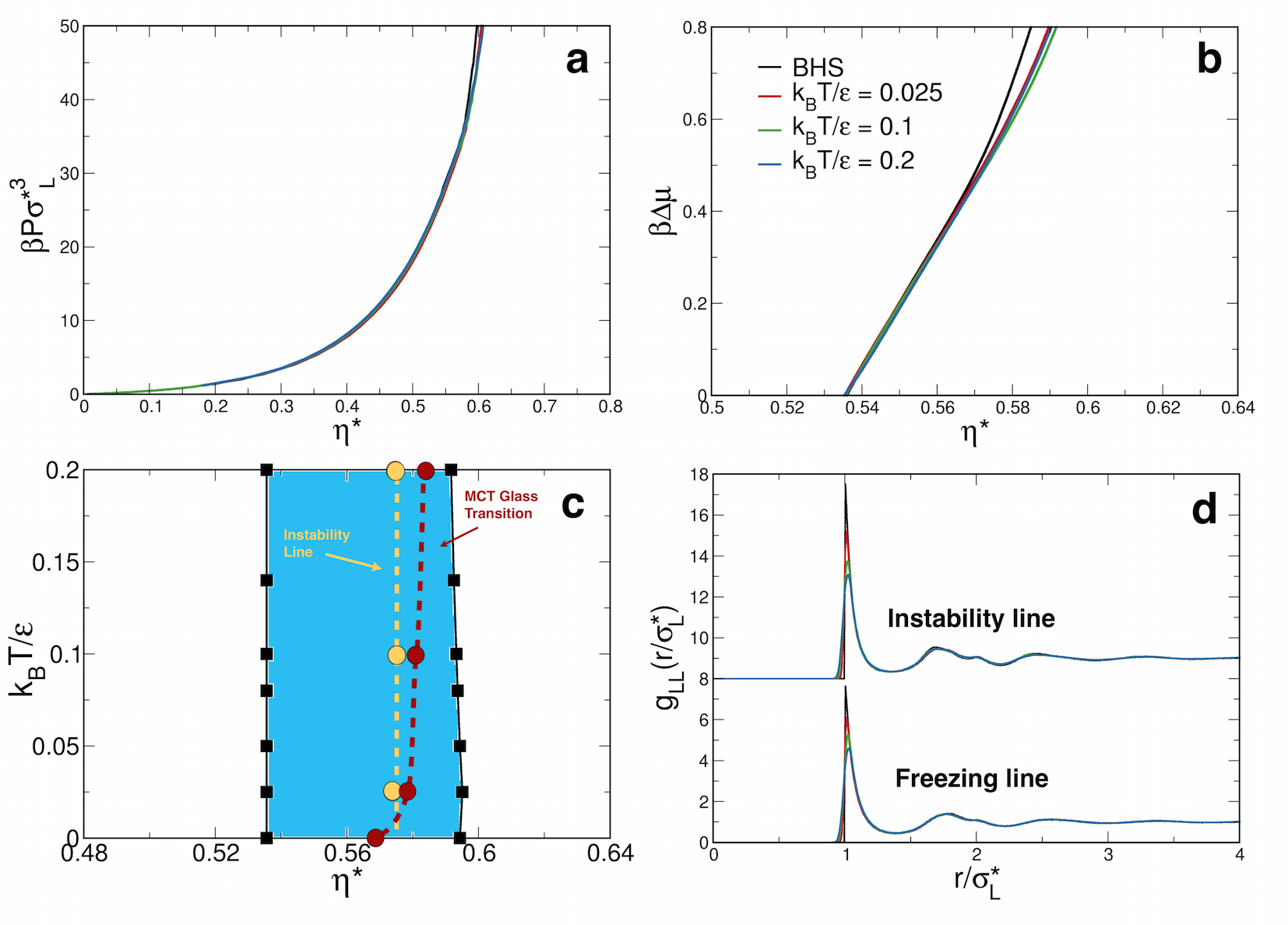}
\caption{a) The reduced pressure $\beta P \sigma_L^{*3}$ and b) the supersaturation $\beta\Delta \mu$ versus effective packing fraction $\eta^*$ for a  binary hard-sphere mixture and a binary WCA mixture for varying temperatures with a diameter ratio $q$ = 0.78 and composition $x_L=1/3$. In c) we show the phase diagram of this binary WCA mixture in the reduced temperature $k_BT/\epsilon$ - $\eta^*$ plane. The yellow circles connected by a vertical dashed line denote the instability line where the fluid in unstable with respect to freezing. The kinetic MCT glass transition points are denoted by the red circles.  d) The pair correlation function $g_{LL}(r)$ for the large spheres as a function of the scaled radial distance $r/\sigma^*_L$ for a binary mixture of WCA spheres at  three different temperatures and for a BHS mixture using MC simulations along the  freezing line ($\beta \Delta \mu=0$) and  along the instability line ($\beta \Delta \mu \simeq 0.53$).}
\label{invariance}
\end{figure}

\noindent \textbf{Thermodynamic invariance.} To investigate the thermodynamic invariance of  our WCA systems, we relate the thermodynamic properties of the WCA systems to those of a reference hard-sphere system.  For this purpose,  we scale the freezing number density of the binary  WCA system  for $q$ = 0.78 at temperature $T^*$ to the binary hard-sphere freezing packing fraction $\eta_\textrm{\scriptsize BHS}^{(f)}$, which allows us to determine an \emph{effective diameter} $\sigma_L^*$ as well as an effective packing fraction $\eta^*$ at each temperature, in a similar way as in Refs.~ \autocite{filion2011simulation,richard2015role}.  For a BHS mixture with a  diameter ratio $q$ = 0.78, the freezing packing fraction is $\eta_\textrm{\scriptsize BHS}^{(f)}$ = 0.5356. In Fig. ~\ref{invariance}c, we present the phase diagram in the temperature $k_BT/\epsilon$ - effective packing fraction $\eta^*$ plane. As the freezing density for all temperatures of the WCA system is scaled to the freezing packing fraction of hard spheres,  the freezing line becomes a vertical line in this representation.  In addition, we find that  the softness of the interactions has only a minor effect on the melting line and the width of the coexistence region. We find that the melting line shifts  slightly to lower packing fractions and that the width of the coexistence region decreases marginally upon increasing the softness of the potential, \emph{i.e.} increasing the reduced temperature $T^*$. 
In addition, we also plot the effective packing fractions corresponding to the state points where the fluid becomes unstable with respect to freezing, i.e. $\beta \Delta \mu \simeq 0.53$. This instability line lies well inside the two-phase coexistence region, and shifts slightly to higher $\eta^*$ with increasing temperature $T^*$. Subsequently, we use the same effective diameter $\sigma_L^*$ to scale the equations of state, $\beta P \sigma_L^{*3}$ versus $\eta^*$ for our WCA systems at different temperatures, and compare them with the equation of state for a BHS mixture with a diameter ratio $q=0.78$. As seen in Fig.~\ref{invariance}a, we find a perfect collapse of the equations of state, demonstrating a thermodynamic invariance for the equations of state for  the WCA systems with temperature. Finally, we  plot the chemical potential difference $\beta \Delta \mu=\beta\mu_{\textrm{\small fluid}}(P)-\beta \mu_{\textrm{\small LP}}(P)$ between the fluid and the Laves phase as a function of the effective packing fraction $\eta^*$ in Fig.~\ref{invariance}b for both the WCA systems at varying temperatures and the BHS mixture. The collapse of the chemical potential difference for the WCA systems with different temperatures and the BHS system yields a fascinating ``universality" in the thermodynamic driving force for nucleation of the LPs,  explaining our observation of  the thermodynamic invariance of the  Gibbs free-energy  barriers,  nucleation rates, and the onset of spinodal-like behaviour as a function of $\beta\Delta\mu$  for different temperatures as described above.

\noindent \textbf{Structural Invariance.}  In order to investigate whether the structure is  also invariant along lines in the phase diagrams identified by equal $\beta \Delta \mu$ values, we measure the pair correlation function $g_{LL}(r)$ for the large spheres as a function of the radial distance expressed in terms of the effective diameter of the large spheres for the WCA systems at the three different temperatures and for the BHS mixture along the (i)  freezing line $\beta \Delta \mu=0$ and (ii) instability line $\beta \Delta \mu \simeq 0.53$, where we made sure that the system remained in the fluid state during our sampling. The results are presented  in Fig. \ref{invariance}d. We find a good collapse of both sets of $g_{LL}(r)$'s as the peak positions coincide, showing a structural invariance of the two-body correlation functions along the freezing and instability lines. Additionally, one observes  that the height of the first peak of the $g_{LL}(r)$ increases and the peak becomes narrower, and thus the $g_{LL}(r)$  becomes more  hard-sphere-like upon lowering the temperature.

The collapse of the phase diagram, equations of state, and pair correlation functions  demonstrate an  invariance of the binary WCA mixtures for varying temperatures along lines of equal  $\beta\Delta\mu$, \emph{i.e.} thermodynamic driving force, in the phase diagram. We thus find that a binary mixture of  soft repulsive spheres can be mapped onto a hard-sphere system in such a  way that the structure and thermodynamics are invariant. However, this invariance does not yet explain why nucleation of LPs is observed in the case of WCA systems and not for a binary hard-sphere mixture. 

\subsection*{Kinetic glass transition} 
\begin{figure}[t!]
\centering
\hspace*{-0.1in}
\includegraphics[width=0.9\textwidth, keepaspectratio, angle=0]{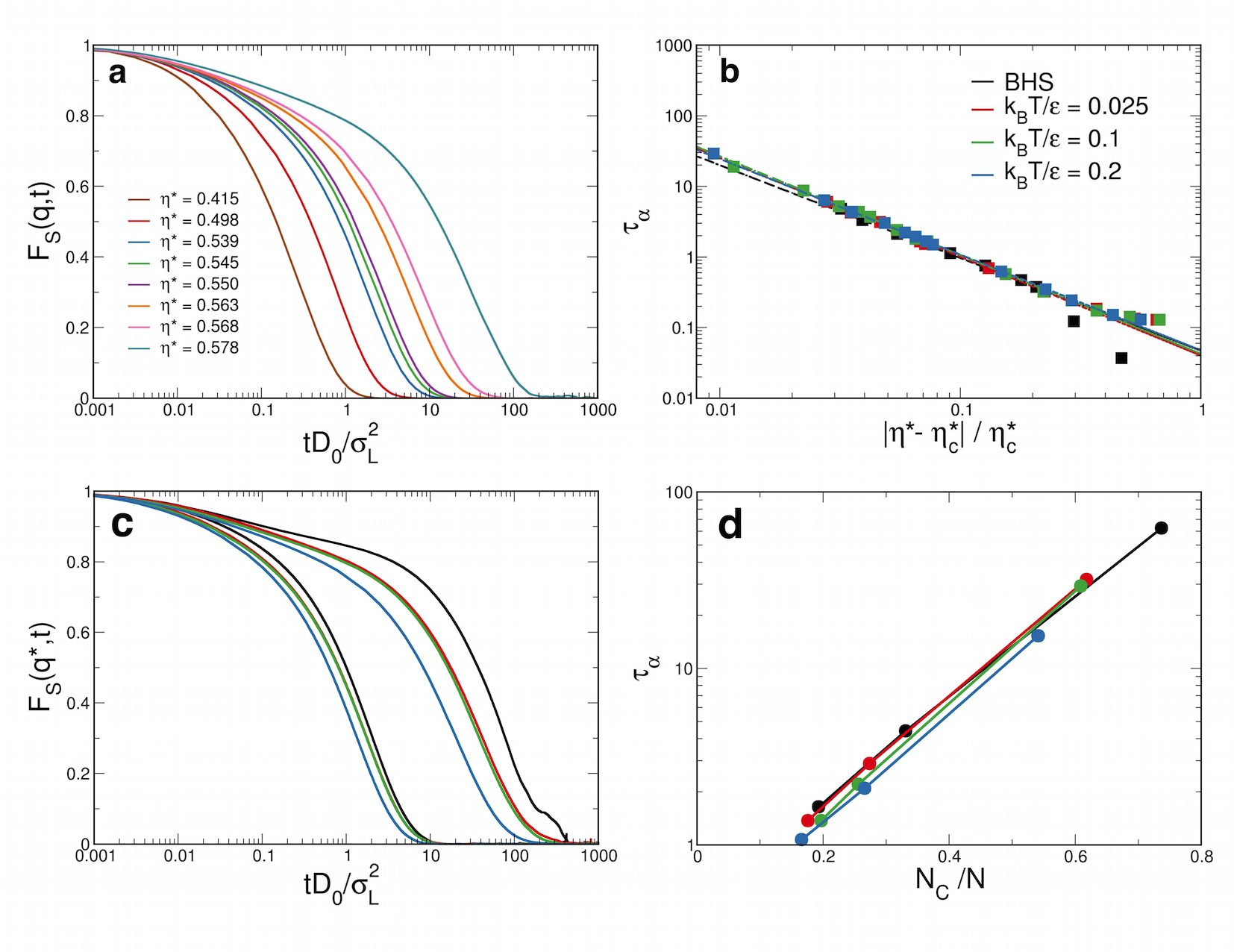}
\caption{a) The self-intermediate scattering function $F_\textrm{\small s}(q,t)$ for the large spheres as a function of time $tD_0/\sigma^2_L$ for a binary WCA mixture  with a  diameter ratio $q=0.78$  at $T^*=0.2$ for varying effective packing fractions $\eta^*$ as  obtained from MC simulations. b)  The structural relaxation time  $\tau_{\alpha}$ as a function of $\mid\eta^*-\eta_c^*\mid/\eta_c^*$  for  a binary mixture of  WCA spheres at $T^*=0.025, 0.1,$ and $0.2$ and a binary  hard-sphere mixture. c) The self-intermediate scattering function $F_\textrm{\small s}(q^*,t)$ for the large spheres  for a binary mixture of WCA spheres at  three different temperatures and for a BHS mixture using MC simulations along the freezing line ($\beta \Delta \mu=0$) and  along the instability line ($\beta \Delta \mu \simeq 0.53$). d) Exponential relation between the relaxation time at three different levels of supercooling ($\beta \Delta \mu = 0$, $\beta \Delta \mu = 0.2$ and $\beta \Delta \mu \simeq 0.53$) and the number fraction of particles belonging to a defective icosahedron.}
\label{MCT}
\end{figure}

To shed light on this counterintuitive result, we investigate the kinetics of the WCA systems as a function of temperature. We already demonstrated above that the degree of five-fold symmetry clusters can be tuned by the softness of the interaction potential. To investigate the effect of five-fold clusters on the kinetics of the system further, we  determine the kinetic glass transition of the WCA systems at varying temperatures. To this end, we calculate  the self-intermediate scattering function $F_s(q,t)= 1/N \sum_{j=1}^{N_L} \left\langle \exp \left\lbrace i {\bf q}.\left[{\bf r}_j(0)-{\bf r}_j(t)\right] \right\rbrace \right\rangle$ at the  wave vector  $q$ = $\vert {\bf q} \vert$ = $2\pi / \sigma_L$ as a function of  time $tD_0/\sigma^2_L$ using MC simulations. 
Exemplarily, we plot  $F_\textrm{\small s}(q,t)$ for a WCA mixture at $T^*=0.2$ in Fig.~\ref{MCT}a for varying $\eta^*$. The dynamics slows down dramatically with increasing $\eta^*$.   At sufficiently high densities, the  structural relaxation time $\tau_{\alpha}$, defined by $F_\textrm{\small s}(q,\tau_{\alpha}) = e^{-1}$, diverges algebraically~\autocite{ni2013pushing}, \emph{i.e.} $\tau_{\alpha}\sim\mid\eta^*-\eta_c^*\mid^{-\gamma}$. Here,  $\eta^*_c$ denotes the critical packing fraction that corresponds to the  kinetic glass transition as  described by mode coupling theory (MCT)~\autocite{gotze1999recent}. We plot the structural relaxation time  $\tau_{\alpha}$ as a function of $\mid\eta^*-\eta_c^*\mid/\eta_c^*$ in Fig.~\ref{MCT}b for our WCA systems at $T^*=0.025, 0.1,$ and $0.2$ and the BHS mixture. We find a perfect collapse of all the data. We fit the relaxation times $\tau_{\alpha}$ using MCT, and list the critical MCT packing fractions $\eta_c^*$ with the corresponding supersaturation $\beta\Delta\mu$, critical exponents $\gamma$,  and the effective diameters $\sigma_L^*$  in  Table~\ref{summarytable} for the WCA systems at varying temperatures, \emph{i.e.} softness of the interaction potential and the BHS mixture. We also plot the critical MCT packing fractions $\eta_c^*$ in Fig. \ref{invariance}c. The results are striking. We clearly observe that the critical MCT effective packing fraction $\eta_c^*$ and corresponding supersaturation $\beta \Delta\mu$ decreases with decreasing softness of the interaction potential. In fact, in the BHS case, the kinetic glass transition, as predicted by MCT, precedes the state point where the fluid becomes unstable with respect to freezing. 

In order to make a further comparison between the investigated systems,  we compare $F_s(q^*,t)$ at the  wave vector  $q^*$ = $\vert {\bf q^*} \vert$ = $2\pi / \sigma^*_L$, for state points along the freezing line $\beta \Delta \mu=0$ and along the instability line  $\beta \Delta \mu \simeq 0.53$ for a BHS system and  WCA systems at $T^*$ = 0.025, 0.1 and 0.2, in Fig. \ref{MCT}c. Not only do we observe a different dynamical behaviour for varying softness, but we also note a remarkable correlation between the relaxation times $\tau_{\alpha}$ and the number fraction of fivefold symmetry defective icosahedron clusters. We display this correlation in Fig. \ref{MCT}d, where we find an exponential relation between the structural relaxation times in  the fluid phase for varying particle softness and different supersaturations $\beta \Delta\mu$, and the fraction of particles belonging to a defective icosahedron. 

To summarise, we find that a BHS mixture gets kinetically arrested at a lower packing fraction than the packing fraction where we expect to find spontaneous nucleation of the LP. However, for a slightly softer interaction potential, a binary WCA mixture at $T^*=0.025$, we find the reverse situation, and hence spontaneous nucleation is observed at a packing fraction that is lower than that of the kinetic glass transition. This finding may explain why LP nucleation is never observed in a binary mixture of hard spheres, and is observed here for a binary WCA system. 

\begin{table}[!t]
\centering
\begin{tabular}{c c c c c} \hline
        \hline
System&$\eta^*_\textrm{\small c}$&$\beta \Delta \mu$&$\gamma$&$\sigma_L^*$\\[0.0 cm]
		\hline	
$k_BT/\epsilon = 0.2$&0.5837&0.673&1.3545&1.0583\\
$k_BT/\epsilon = 0.1$&0.5816&0.620&1.3984&1.0764\\
$k_BT/\epsilon = 0.025$&0.5792&0.604&1.3993&1.1009\\
BHS&0.5681&0.452&1.3140&1.0000\\ 
\hline
\end{tabular}
\caption{\label{summarytable} The critical MCT effective packing fraction $\eta_c^*$ corresponding to the kinetic glass transition for a binary mixture of WCA spheres at varying temperatures and for a BHS mixture, all at a diameter ratio $q=0.78$, the corresponding supersaturation level $\beta\Delta\mu$, the critical exponents $\gamma$ of the MCT fits,  and the effective large-sphere diameters $\sigma_L^*$. }
\end{table}

\section*{Discussion}
In 2007, a  novel self-assembly route towards a photonic bandgap material was proposed in which the diamond and pyrochlore structure are self-assembled  from a binary mixture of colloidal hard spheres into a closely packed MgCu$_2$ Laves phase~\autocite{hynninen2007self}. Despite numerous efforts, spontaneous crystallization of the LPs has never been observed  in simulations of BHS mixtures or in experiments on micron-sized colloidal hard spheres, casting doubts on the thermodynamic stability of these crystal structures in binary hard spheres. Recent MC simulations have shown, however, that by introducing size polydispersity, either in a static or dynamic way, and by using unphysical particle swap  moves, LPs may be nucleated from a dense hard-sphere fluid \autocite{lindquist2018communication,bommineni2019complex}. Alternatively, to alleviate problems with the degeneracy of the three competing LPs and with the metastability of the MgCu$_2$  w.r.t. the MgZn$_2$ phase, one may resort to another self-assembly route in which  the   MgCu$_2$ phase, stable in the present system, is formed from  a binary mixture of colloidal spheres and preassembled tetrahedral clusters of spheres as shown both in simulations \autocite{avvisati2017fabrication} and experiments using DNA-mediated interactions  \autocite{ducrot2017colloidal}. To better understand why the nucleation of  LP is severely hampered in a binary fluid of hard spheres, we investigated the degree of fivefold symmetry in the binary fluid phase as the presence of fivefold symmetry structures may suppress nucleation. In order to study the effect of softness of the interaction potential, we measured the number fraction of three significant representatives of the fivefold symmetry structures  in a binary fluid of WCA spheres at varying temperatures $k_BT/\epsilon$ thereby altering the softness of the interaction potential. In the limit of $k_BT/\epsilon \rightarrow 0$, this system reduces to the binary hard-sphere system.  Surprisingly, we found  that particle softness significantly reduces the degree of fivefold symmetry in the binary fluid phase.  To investigate the repercussions of this finding on LP nucleation, we subsequently performed simulations with a crystalline seed to measure the nucleation barrier  and nucleation rate for the three LP types and for varying temperatures, \emph{i.e.} degrees of particle softness. These results enabled us to study, for the first time, spontaneous nucleation of the LPs in simulations of  nearly hard spheres. We thus find that the seeding approach is versatile and robust ~\autocite{espinosa2016seeding,espinosa2017lattice} --- it not only enables one to determine the nucleation barrier and nucleation rate, but also locate the regime in the phase diagram where spontaneous nucleation may occur and provides information on how a crystal nucleus grows and melts. 
Our observation of spontaneous nucleation of the LP in a system of soft spheres is important and  intriguing for two reasons. On the one hand, our simulations provide evidence that the LP  is  stable  in the phase diagram of such a binary mixture, as predicted theoretically more than a decade ago \autocite{hynninen2007self}.   On the other hand, it immediately begs the question why LP nucleation has never been seen  in simulations of BHS mixtures or in experiments on micron-sized colloidal hard spheres despite numerous attempts by many research groups, whereas it nucleates spontaneously with a tiny degree of particle softness.  To address this question, we studied the role of softness in the interaction potential on the structure, phase behaviour, and dynamics of the LPs, and found that a system of soft repulsive spheres can be mapped onto a binary hard-sphere system in such a way that the structure and thermodynamics are invariant in reduced units for varying softness of the interaction potential. However,  the invariance of the nucleation barrier and nucleation rate as a function of  supersaturation for varying softness of the potential seems to be at odds with the observation of LP nucleation in WCA systems and the absence of it in binary hard spheres.  In order to shed light on this counterintuitive result, we determined the kinetic glass transition by fitting the structural relaxation times as obtained from the self-intermediate scattering functions with an MCT fit for the various WCA systems. Surprisingly, we found that the packing fraction corresponding to the kinetic glass transition  strongly depends on the softness of the particle interactions, which in its turn affects the presence of fivefold symmetry clusters in the supersaturated fluid phase. We thus find that crystallization can be enhanced by tuning the softness of the particle interactions, either by charge, ligands, or a stabilizer, in simulations or experiments. This finding is indeed consistent with  the experimental observations of the LPs as they all seem to involve particles interacting  with  (slightly) soft repulsive interactions~\autocite{shevchenko2006structural,evers2010entropy, yoshimura1983order,hasaka1984structure,ma1994preparation,gauthier2004phase,cabane2016hiding,schaertl2018formation}. Moreover, introducing a small degree of softness in the particle interactions can be exploited in a wealth of other crystallization studies. For instance, there are still many open questions on how and why binary crystal phases nucleate. A systematic study of binary nucleation has been hampered so far by either slow dynamics or by finding the right regime in the phase diagram where nucleation may occur.  
Finally, we note that the structure as characterized by the two-body correlation functions as well as the thermodynamics which is predominantly determined by also two-body correlations is invariant in reduced units for varying softness of the pair potential. However, the structural relaxation time and the kinetics depend strongly on the presence of fivefold structures, and  thus on higher-body correlations. We hope that this finding will inspire the development of new theories for predicting the kinetic glass transition that take into account higher-body correlations.

\section*{Methods}
\subsection*{Computer simulations}
We investigate the crystallization of the Laves phases in nearly hard spheres by simulating a binary mixture  of $N_L$ large ($L$)  and $N_S$ small ($S$) spheres interacting with a Weeks-Chandler-Andersen (WCA) potential $u_{\alpha\beta}(r_{ij})$   between  species $\alpha=L,S$ and $\beta=L,S$ \autocite{weeks1971role} 
\begin{align*}
u_{\alpha\beta} \left(r_{ij}\right) &= 4 \epsilon_{\alpha\beta} \left\lbrack \left(\frac{\sigma_{\alpha\beta}}{r_{ij}} \right)^{12} - \left(\frac{\sigma_{\alpha\beta}}{r_{ij}} \right)^6 + \frac{1}{4} \right\rbrack & r_{ij}<2^{1/6}\sigma_{\alpha\beta}\nonumber\\
&= 0 &r_{ij}\geq2^{1/6}\sigma_{\alpha\beta}, 
\label{wca}
\end{align*}
where $r_{ij}=|{\bf r_i}-{\bf r_j}|$ denotes the center-of-mass distance between particle $i$ and $j$, $r_i$ the position of particle $i$, and $\epsilon_{LL} = \epsilon_{SS} = \epsilon_{LS} = \epsilon_{SL} = \epsilon$ the interaction strength. We set the diameter ratio $q = \sigma_{S}/\sigma_{L} = 0.78$ with $\sigma_{\alpha}$ the diameter of species $\alpha$, and use $\sigma_{\alpha\beta} = \left(\sigma_{\alpha} + \sigma_{\beta}\right)/2$.  The softness of the potential can be tuned by changing the reduced temperature $T^*=k_BT/\epsilon$ with  $k_B$ Boltzmann's constant and $T$ the temperature.

\noindent In the determination of the kinetic glass transition and the degree of fivefold symmetry clusters at different supersaturations, MC simulations were performed on $N = N_L+N_S =$ 1200 particles (WCA spheres and binary hard spheres) with composition $x_L = N_L/(N_L+N_S) = $ 1/3, in the \emph{NVT} ensemble involving standard single particle translation moves. The cluster concentrations are averaged over 100 independent snapshots. 
Spontaneous nucleation of LPs was observed in MD simulations performed using HOOMD-blue (Highly Optimized Object-oriented Many-particle Dynamics) \autocite{Anderson2008,Glaser2015} in the $NPT$ ensemble on 1536 WCA spheres at composition $x_L = $ 1/3. The temperature $T$ and isotropic pressure $P$ are kept constant \textit{via} the Martyna-Tobias-Klein (MTK) \autocite{martyna1994constant} integrator, with the thermostat and barostat coupling constants $\tau_{\tiny T}$ = 1.0 $\tau_{\scriptsize MD}$ and $\tau_{\tiny P}$ = 1.0 $\tau_{\scriptsize MD}$ respectively, where $\tau_{\scriptsize MD} = \sigma_{\scriptsize L}\sqrt{m/\epsilon}$ is the MD time unit. The time step is set to $\Delta t = 0.004\tau_{\scriptsize MD}$ and the simulations are run for 10$^{9} \tau_{\scriptsize MD}$ time steps, unless otherwise specified. The simulation box is cubic and periodic boundary conditions are applied in all directions. 

\noindent \textbf{Seeding approach.} The nucleation free-energy barrier heights and nucleation rates were calculated using a crystal seeding approach involving a two-step equilibration process as described in the main text, \emph{via} MC simulations in the $NPT$ ensemble involving isotropic volume scale moves in addition to the particle translation moves. The initial configurations were seeded with all three LP types MgCu$_2$, MgZn$_2$ and MgNi$_2$, surrounded by disordered particles, at overall composition $x_L = $ 1/3. The data points on the free-energy barrier height and nucleation rate profiles are obtained from simulations with five different seed sizes ($N_{\scriptsize seed}$): (i) $N_{\scriptsize seed}$ = 96 (all three LPs) and total system size $N$ = 4140, (ii) $N_{\scriptsize seed}$ = 192 (MgCu$_2$), 384 (MgZn$_2$, MgNi$_2$) and $N$ = 4160, (iii) $N_{\scriptsize seed}$ = 648 (MgCu$_2$), 1080 (MgZn$_2$), 720 (MgNi$_2$) and $N$ = 8100, (iv) $N_{\scriptsize seed}$ = 1536 (MgCu$_2$), 1728 (MgZn$_2$), 1440 (MgNi$_2$) and $N$ = 12500, and (v) $N_{\scriptsize seed}$ = 3000 (MgCu$_2$), 2688 (MgZn$_2$, MgNi$_2$) and $N$ = 17000 particles, respectively. The critical pressures and attachment rates for the different seed sizes were obtained from MD simulations in the \emph{NPT} ensemble, where the simulations were initialised with the MC-equilibrated configurations. The details of the MD simulations are described above. 

\subsection*{Free energy calculations}
The equilibrium phase diagram of the WCA mixture is calculated by determining the free energies of the binary fluid at composition $x_L$ = 1/3 and the three LPs - MgCu$_2$, MgZn$_2$, and MgNi$_2$. The Helmholtz free energy per particle $f=
F/N$ as a function of density $\rho$ for all these phases is calculated using thermodynamic integration of the equations of state
\begin{equation}
\beta f\left(\rho\right) = \beta f\left(\rho_0 \right) + \int_{\rho_0}^{\rho} d\rho{'}   \frac{\beta P(\rho{'})}{\rho'\\^{2}}, 
\label{eosint}
\end{equation}
where $\rho=N/V$ is the density with $N$ the number of particles and $V$ the volume of the system, $f\left(\rho_0\right)$ denotes the Helmholtz free energy per particle for the reference density $\rho_0$, $\beta= 1/k_B T$ is the inverse temperature, and $P$ is the pressure. The equations of state for the binary fluid and the binary LPs are calculated using MC simulations in the \emph{NPT} ensemble. Isotropic volume change moves are used for the fluid phase and the cubic MgCu$_2$ phase while anisotropic volume change moves are used for the hexagonal MgZn$_2$ and MgNi$_2$ phases. We use the ideal gas as a reference state for the binary fluid phase. 
For the LPs, we employ the Frenkel-Ladd method to calculate the Helmholtz free energy at a reference density $\rho_0$ using MC simulations in the $NVT$ ensemble. In the Frenkel-Ladd method, we start from an Einstein crystal, where the particles are coupled via harmonic springs with a dimensionless spring constant $\lambda$ to the ideal positions  of the crystal structure under consideration.  We then construct a reversible path from the crystal of interest to the Einstein crystal using the auxiliary potential energy function 
\begin{equation}
\beta U_{Ein}({\bf r}^N;\lambda) = \beta U({\bf r}_0^N) + (1-\frac{\lambda}{\lambda_{max}})\lbrack\beta U({\bf r}^N)- \beta U({\bf r}_0^N)\rbrack +  \lambda \sum_{i=1}^N \frac {({\bf r}_i -{\bf  r}_{0,i})^2}{\sigma_L^2}, 
\end{equation}
where $U({\bf r}^N)= \sum_{i<j}^N u(r_{ij})$ is the potential energy of the system due to the interparticle interactions, ${\bf r}_{0,i}$ represents the ideal lattice position of particle $i$, and $\lambda$ is the dimensionless spring constant, which ranges from 0 to a value $\lambda_{max}$. At $\lambda_{max}$, the particles are so strongly tied to their respective lattice sites, that the systems reduces to an Einstein crystal of non-interacting particles, whereas $\lambda=0$ corresponds to the interacting system of interest for which we want to compute the free energy \autocite{polson2000finite,vega2008determination}. 
The Helmholtz free energy of the crystal $\beta f(\rho)$ can be approximated to that of the Einstein crystal using \autocite{frenkel2001understanding}:
\begin{eqnarray}
\beta f(\rho) = \beta f_{Ein}(\lambda_{max}) - \frac{1}{N} \int \limits_{\lambda = 0}^{\lambda_{max}} d\lambda \left\langle\frac{\partial \beta U_{Ein}({\bf r}^N;\lambda)}{\partial \lambda}\right\rangle_{\lambda}^{CM}
\label{fe_expression}
\end{eqnarray}
where  $\langle \cdots \rangle_{\lambda}^{CM}$ denotes that the ensemble average is sampled for a solid with a fixed center of mass using the Boltzmann factor $\exp[-\beta U_{Ein}({\bf r}^N;\lambda)]$, and $f_{Ein}(\lambda_{max}) $ denotes  the free energy per particle of an ideal Einstein crystal given by 
\begin{eqnarray}
\beta f_{Ein}(\lambda_{max}) = \frac{\beta U({\bf r}_0^N)}{N} + \frac{3(N-1)}{2N} \ln \left(\frac{\lambda_{max}}{\pi }\right)+\frac{1}{N} \ln \left(\frac{N}{V}\Lambda^3\right)-\frac{3}{2N}\ln(N),
\label{fe_ein}
\end{eqnarray}
where $U({\bf r}_0^N)$ is the potential energy when all particles are at their ideal lattice positions, $\Lambda$ is the thermal wavelength. 
We note that it is convenient to rewrite the integral in Eq.~\ref{fe_expression} as
\begin{eqnarray}
\frac{1}{N} \hspace{-0.1in}\int \limits_{\ln c}^{\ln(\lambda_{max} + c)} \hspace{-0.2in}\left(\lambda + c\right) \left\langle \sum_{i=1}^N \frac{\left({\bf r}_i -{\bf  r}_{0,i}\right)^2}{\sigma_L^2} -\frac{1}{\lambda_{max}}\lbrack\beta U({\bf r}^N)- \beta U({\bf r}_0^N)\rbrack\right\rangle_{\lambda}^{CM} d\left[\ln\left(\lambda + c\right)\right], 
\label{thermoint}
\end{eqnarray}
with
\begin{equation}
c = \frac{1}{\left\langle \sum_{i=1}^N \left({\bf r}_i - {\bf r}_{0,i}\right)^2/\sigma_L^2  -\frac{1}{\lambda_{max}}\lbrack\beta U({\bf r}^N)- \beta U({\bf r}_0^N)\rbrack\right\rangle_{\lambda = 0}^{CM}}.
\end{equation} 
The integral in Eq. \ref{thermoint} is calculated numerically using a 40 or 60 point Gauss-Legendre quadrature, yielding the LP free energy at the reference density. We find that the MgCu$_2$ and MgNi$_2$ LPs are metastable compared to the MgZn$_2$ LP (see Supplementary Fig. 1). Subsequently, the LP-fluid coexistence densities for different temperatures $T^*$ are calculated by employing a common-tangent construction on the fluid and LP free-energy density curves in the $\beta f \rho$ -- $\rho$ plane. The supersaturation $\beta \Delta \mu=\beta\mu_{\textrm{\small fluid}}(P)-\beta \mu_{\textrm{\small LP}}(P)$ signifying the chemical potential difference  between the supersaturated fluid and the stable LP at pressure $P$, can be determined by  employing the Gibbs-Duhem relation 
\noindent $ \int_{\mu (P_{coex})}^{\mu (P)} d \mu' = \int_{P_{coex}}^{P} \frac{1}{\rho(P')} dP'$ with $P_{coex}$ and $\mu_{coex}$ the bulk pressure and chemical potential at the fluid-LP coexistence. 

\subsection*{LP cluster identification}
In order to study the nucleation of the Laves phases, we require a criterion that distinguishes crystalline clusters with the Laves phase (LP) symmetry from the binary fluid phase. 
To this end, we first make a distinction between particles that have a solid-like environment with an LP-like symmetry and a fluid-like environment. We use the local bond orientational order parameters $q^{\alpha}_{l,m}(i)$  to determine the symmetry of the local environment of particle $i$ with identity $\alpha(i) \in L,S$ \autocite{steinhardt1983bond}
\begin{equation} 
q^{\alpha}_{l,m}(i) = \frac{1}{N_{b}(i)} \sum_{j=1}^{N_{b}(i)} Y_{l,m}\left( \theta_{i,j}, \phi_{i,j} \right),
\end{equation}
where $N_{b}(i)$ represents the number of all neighbours of particle $i$ with identity $\beta(j)$, $Y_{l,m}(\theta,\phi)$ are the spherical harmonics for $m$ ranging from $[-l,l]$, and $\theta_{i,j}$ and $\phi_{i,j}$ are the polar and azimuthal angles of the center-of-mass distance vector ${\bf r}_{ij}={\bf r}_j-{\bf r}_i$ with ${\bf r}_i$ the position of particle $i$. The neighbours of particle $i$ with identity $\alpha(i)$ include all particles $j$ with identity $\beta(j)$ that lie within a radial distance $r_c$ of particle $i$, and we set $r_c$ equal to the distance corresponding to the first minimum of the respective    radial distribution function.  
Subsequently, we calculate the dot product $d^{\alpha\beta}_l(i,j)$ for each particle pair $i$ with identity $\alpha(i)$ and $j$ with identity $\beta(j)$  
\begin{equation}
d_{l,\alpha\beta}(i,j) =\frac{\sum\limits_{m=-l}^l q^{\alpha}_{l,m}(i) q^{\beta}*_{l,m}(j)}{\left( \sum_{m=-l}^{l} \vert q^{\alpha}_{l,m}(i) \vert ^{2} \right)^{1/2}   \left( \sum_{m=-l}^{l} \vert q^{\beta}_{l,m}(j) \vert ^{2} \right)^{1/2}}.
\label{dot}
\end{equation}
We note that the dot product is symmetric in $\alpha$ and $\beta$, i.e., $d_{l,\alpha\beta}(i,j)=d_{l,\beta\alpha}(i,j)$. We choose a symmetry index $l=6$.  Only in the case $\alpha = \beta = L$, we employ the \emph{average} bond order parameter $\overline{q}^{\alpha}_{l,m}(i)$ as introduced in Ref. \autocite{lechner2008accurate}
\begin{equation}
\overline{q}^{L}_{l,m}(i) = \frac{1}{N_{b}(i)+1}\sum_{k=0}^{N_{b}(i)} q^{L}_{l,m}(k), 
\label{lech}
\end{equation}
where $N_{b}(i)$ denotes the number of ($L$) neighbours of ($L$) particle $i$ and $k = 0$ represents the particle $i$ itself, and we use this average bond order parameter in the dot-product expression of Eq. \ref{dot}. 
The bond between particles $i$ and $j$ is classified as a \emph{solid} bond if the dot product $d_{l,ij}$ lies in between a lower and a upper threshold value denoted by $d^{\downarrow}_{\alpha\beta}$ and $d^{\uparrow}_{\alpha\beta}$. 
In order to discriminate   crystalline clusters with the Laves phase symmetry from the fluid phase, we use the following cut-off values, as determined from the intersections of the dot product distributions of the fluid and  Laves phases (see Supplementary  Fig. 2) for the three different species correlations. For the (i) large-large correlation, we employ  $\overline{d}_{6,LL} >  d^{\downarrow}_{LL}=0.88$, (ii) small-small correlation $d^{\downarrow}_{SS}=-0.23 < d_{6,SS} <  d^{\uparrow}_{SS}=-0.04$ and (iii) small-large  $d^{\downarrow}_{LS}=-0.46 < d_\textrm{6,LS} <  d^{\uparrow}_{LS}=-0.1$. Using these threshold values, we define a particle $i$ with identity $\alpha$ as \emph{crystalline} if the number of solid bonds  satisfies $\xi^{\alpha}(i)> \xi^{\alpha}_c$ where 
\begin{equation}
\xi^{\alpha}(i) = \sum_{j=1}^{N_{b}(i)}H(d_{l,\alpha\beta}(i,j)-d^{\downarrow}_{\alpha\beta}) - H(d_{l,\alpha\beta}(i,j)-d^{\uparrow}_{\alpha\beta})
\label{heaviside}
\end{equation}
and $H$ is the Heaviside step function.  We employ the threshold values $\xi^L_c = 10$ and $\xi^S_c = 9$ for the large and small species, respectively. The above criteria are sufficient to distinguish the crystalline particles with LP-like symmetry from the surrounding fluid.

In order to investigate the effect of fivefold symmetry clusters on crystallization, we require an algorithm that is capable of successfully finding different topological clusters in a metastable fluid. To this aim, we employ the Topological Cluster Classification (TCC) \autocite{malins2013identification} for varying softness of the interparticle potential. The algorithm is used regardless of the species of the particles forming the clusters and the bonds between particles are detected using a modified Voronoi construction method. The free parameter $f_c$, controlling the amount of asymmetry that a four-membered ring can show before being identified as two three-membered rings, is set to $0.82$.

\printbibliography

\subsection*{Acknowledgements} 
T.D. and M.D. acknowledge financial support from the Industrial Partnership Programme, ``Computational Sciences for Energy Research" (Grant no. 13CSER025), of the Netherlands Organization for Scientific Research (NWO), which was co-financed by Shell Global Solutions International B.V. G.M.C. was also financially supported by NWO (Grant no. 16DDS003). 

\subsection*{Author contributions}
M.D. initiated the project on the role of softness in LP nucleation in binary nearly hard spheres, and supervised T.D. and G.M.C. T.D. performed free-energy calculations and computed the thermodynamic phase diagram. G.M.C.
determined the cluster criterion that distinguishes crystalline clusters with the Laves phase symmetry, and calculated the degree of fivefold symmetry in the binary fluid as a function of softness and supersaturation. G.M.C. and T.D. determined the nucleation free-energy barrier, nucleation rate, kinetic glass transition, and established thermodynamic and structural invariance for different degrees of softness, using MC and MD simulations respectively. All authors co-wrote the manuscript, analysed, and discussed the interpretation of the results.

\subsection*{Competing interests}
The authors declare no competing interests.

\end{document}